\documentclass[nofootinbib,twocolumn,prd,showpacs]{revtex4}

\usepackage[dvips]{graphicx}
\usepackage{amsmath,amssymb,amsopn,bm}
\usepackage{color}
\usepackage{epstopdf}
\usepackage[normalem]{ulem}
\usepackage{subfigure}
\usepackage{verbatim}

\newcommand{\avg}[1]{\left< #1 \right>} % for average
\newcommand{\ket}[1]{\left| #1 \right>} % for Dirac bras
\newcommand{\bra}[1]{\left< #1 \right|} % for Dirac kets
\newcommand{\M}{\ensuremath{\mathcal{M}}} % for Borel mass
\begin{document}

\title{Quantitative sum rule analysis of low-temperature spectral functions}

\author{Nathan P. M. Holt}
\email{nathan.holt@physics.tamu.edu}
\affiliation{Cyclotron Institute and Department of Physics\,\&\,Astronomy,
Texas A{\&}M University, College Station, Texas 77843-3366, USA }

\author{Paul M. Hohler}
\email{pmhohler@comp.tamu.edu}
\affiliation{Cyclotron Institute and Department of Physics\,\&\,Astronomy,
Texas A{\&}M University, College Station, Texas 77843-3366, USA }

\author{Ralf Rapp}
\email{rapp@comp.tamu.edu}
\affiliation{Cyclotron Institute and Department of Physics\,\&\,Astronomy,
Texas A{\&}M University, College Station, Texas 77843-3366, USA }

\date{\today}

\begin{abstract}
We analyze QCD and Weinberg-type sum rules in a low-temperature pion gas
using vector and axial-vector spectral functions  following from the
model-independent chiral-mixing scheme. Toward this end we employ recently
constructed vacuum spectral functions with ground and first-excited states
in both channels and a universal perturbative continuum; they quantitatively
describe hadronic $\tau$-decay data and satisfy vacuum sum rules.
These features facilitate the implementation of chiral mixing without
further assumptions, and lead to in-medium spectral functions which exhibit
a mutual tendency of compensating resonance and dip structures, suggestive
for an approach toward structureless distributions. In the sum rule analysis,
we account for pion mass corrections, which turn out to be significant.
While the Weinberg sum rules remain satisfied even at high temperatures,
the numerical evaluation of the QCD sum rules for vector and axial-vector
channels reveals significant deviations setting in for temperatures beyond
$\sim$140\,MeV, suggestive of additional physics beyond low-energy chiral pion
dynamics.
\end{abstract}

\maketitle

%%%%%%%%%%%%%%%%%%%%%%%%%
\section{Introduction}
\label{sec:intro}
%%%%%%%%%%%%%%%%%%%%%%%%%

The Lagrangian of quantum chromodynamics (QCD) is invariant under the
transformations described by the chiral symmetry group
$SU_{L}(N_{f}) \times SU_{R}(N_{f})$, where $N_{f}$ is the number of light-quark
flavors \cite{Shuryak:2004book,Ioffe:2010zz,Friman:2011zz,Satz:2012zz}. In
vacuum and at low temperatures, this symmetry is spontaneously broken by a
nonzero expectation value of the quark condensate, $\bra{0}\bar{q}q\ket{0}$.
For isovector hadronic
excitations, this manifests itself through a large splitting of the masses
of the chiral partner mesons $\rho(770)$ and $a_{1}(1260)$~\cite{PDG:2012}.

Chiral symmetry is believed to be restored at high temperatures and baryon
densities. As with any symmetry, its breaking and restoration can be
described by order parameters. For the chiral phase transition of QCD,
the canonical order parameter is the quark condensate. Lattice-QCD
computations have found that the quark condensate ``melts'' with increasing
temperature~\cite{Borsanyi:2010bp,Bazavov:2011nk} until its disappearance
signals the restoration of chiral symmetry. The gradual dissolution of the
quark condensate is expected to be accompanied by an emerging degeneracy
of chiral partner hadrons. Thus, the $\rho$ and $a_1$ masses, or more
generally, the pertinent vector and axial-vector spectral functions,
should approach each other with rising temperature and finally become
degenerate when chiral symmetry is restored.

The interplay between the temperature dependence of spectral functions and
condensates can be analyzed using finite-temperature sum rules. These
relate integrals over spectral distributions to an operator
product expansion (OPE) which is expressed in terms of low-energy
condensates. The most relevant sum rules for the present work are the QCD
sum rules (QCDSRs)~\cite{Shifman:1978bx,Shifman:1978by} and the
Weinberg-type sum rules (WSRs)~\cite{Weinberg:1967kj,Das:1967ek,Kapusta:1994}.
The former are channel specific, i.e., they differ for the vector
and axial-vector channels, and they contain both chirally breaking
and symmetric operators. The latter relate moments of the difference between
the vector and axial-vector spectral functions to chiral order parameters.
Both sets of sum rules provide useful tools from which either the
temperature dependence of the spectral functions can be constrained, or
by which models of temperature dependence of the spectral functions and
condensates can be tested~\cite{Friman:2011zz}.

A low-temperature medium at vanishing chemical potential can be represented
by a dilute gas of thermally excited pions. In Ref.~\cite{Dey:1990ba}, it
was shown that the pertinent in-medium vector and axial-vector spectral
distributions can be deduced model-independently as linear combinations of
the vacuum distributions, where the strength of the ``mixing" is determined
by thermal pion exchange
(cf.~also~\cite{Steele:1996su,Chanfray:1998hr,Krippa:1997ss,Harada:2008hj}).
It was further noted that the mixing effect straightforwardly satisfies the
WSRs, which has been verified subsequently~\cite{Eletsky:1992xd,Kapusta:1994}.
Conversely, the
finite-temperature QCDSRs have not received much attention in the context
of chiral mixing. Rather, the (low-) temperature dependence of the
condensates has been evaluated and subsequently used to infer properties
of vector and axial-vector spectral
functions~\cite{Hatsuda:1992bv,Hofmann:1999nn,Zschocke:2002mn}.
More recently, chirally mixed spectral functions have been implemented
within so-called finite-energy sum rules (FESRs) which are obtained from
moments of the QCDSRs~\cite{Marco:2001dh,Kwon:2010fw}.

In the present work, we revisit the use of chirally mixed spectral
functions in both WSRs and QCDSRs utilizing updated vacuum spectral
functions~\cite{Hohler:2012xd} and numerically evaluating the consequences
of finite pion mass corrections. Certain features of the vacuum spectral
functions, such as their degenerate continuum in vector and axialvector
channels and quantitative agreement with $\tau$-decay
data~\cite{Barate:1998uf,Ackerstaff:1998yj}, as well QCDSRs and WSRs,
will turn out to be important in a quantitative low-temperature analysis.

The remainder of this paper is organized as follows. In
Sec.~\ref{sec:sumrules}, we recollect vacuum and finite-temperature sum rules
and discuss the leading temperature dependence and pion-mass corrections
of the OPE coefficients. In Sec.~\ref{sec:FTSF}, we construct the
finite-temperature vector and
axial-vector spectral functions and discuss their characteristics. In
Sec.~\ref{sec:analysis}, we quantitatively evaluate the in-medium
WSRs and QCDSRs employing the chirally mixed spectral functions.
We conclude in Sec.~\ref{sec:conclusion}.

%%%%%%%%%%%%%%%%%%%%%%%%%%%%%%%%%
\section{QCD and Weinberg-type Sum Rules}
\label{sec:sumrules}
%%%%%%%%%%%%%%%%%%%%%%%%%%%%%%%%

We begin by briefly reviewing the vacuum QCDSRs and WSRs (Sec.~\ref{ssec:vac})
and their extension to finite temperature (Secs.~\ref{ssec:QCDSR} and
\ref{ssec:WSR}, respectively). More detailed accounts for the vacuum
case can be found in the original
works~\cite{Shifman:1978bx,Shifman:1978by,Weinberg:1967kj,Das:1967ek,Kapusta:1994}
and at finite temperature in
Refs.~\cite{Hatsuda:1992bv,Kapusta:1994,Zschocke:2002mn,Leupold:1998bt}.

%%%%%%%%%%%%%%%%%%%%%%%%%%%%%%%%%
\subsection{Vacuum sum rules}
\label{ssec:vac}
%%%%%%%%%%%%%%%%%%%%%%%%%%%%%%%%%%
QCD sum rules utilize a dispersion relation to relate a spectral function to an
operator product expansion.  A commonly used form (for operators with dimension less than 6) reads
\begin{equation}
\label{eqn:SR}
\frac{1}{\M^2}\! \int\! \frac{\rho(s)}{s} \, e^{-s/ \M^2} \,\mathrm{d}s = c_0  + \frac{c_1}{\M^2} + \frac{c_2}{\M^4} + \frac{c_3}{\M^6} + \ldots,
\end{equation}
where $\rho(s)$ is the spectral function of a given channel -- $\rho_{V}(s)$
in the vector and
$\bar{\rho}_A(s) = \rho_A(s) + f_\pi^2 \, s \, \delta(s-m_\pi^2)$
in the axial-vector channel where the bar indicates the inclusion of the pion pole.
The ellipsis designates terms associated with operators of higher dimension.
A Borel transformation has traded the space-like four-momentum $q^2=-Q^2$
for the Borel mass $\M^2$. The coefficients appearing on the right-hand side
involve various condensates. For each condensate, we work to leading order
in the quark mass, $m_q$; that is, we consider the effects of each condensate
but not any finite quark mass correction. In the vector and axial-vector
channels, one has
\begin{eqnarray}
c_{0}^{V}  =  c_{0}^{A}
&=& \frac{1}{8\pi^{2}}\left(1+\frac{\alpha_{s}}{\pi}\right) \ ,
\nonumber\\
c_1^V  =  c_1^A &=& -\frac{3}{8} (m_u^2 + m_d^2) \approx 0  \ ,
\nonumber \\
c_{2}^{V}  =  c_{2}^{A}
&=& \frac{1}{24} \langle \frac{\alpha _s}{\pi} G _{\mu \nu}^2
\rangle + m_q \langle \bar{q} q \rangle \ , \\
c_3^V  &=&  -\frac{56}{81} \pi \alpha_s \langle \mathcal{O}_4^V \rangle \ ,
\nonumber\\
c_3^A  &=&  \frac{88}{81} \pi \alpha_s \langle \mathcal{O}_4^A \rangle
\nonumber \ .
 \end{eqnarray}
The gluon and quark condensate are denoted by
$\langle \frac{\alpha _s}{\pi}  G^2 _{\mu \nu}  \rangle$ and
$\langle \bar{q} q \rangle$,
respectively, where $m_q$ is the average current light-quark mass;
$\langle \mathcal{O}_4^V \rangle$ and $\langle \mathcal{O}_4^A \rangle$
refer to the vector and axial-vector four-quark condensates. Note that
the only difference between the vector and axial-vector channels is due
to the chirally odd $c_3$ terms. The four-quark condensates may be written
in a factorized form,
\begin{equation}
\langle\mathcal{O}_4^{V,A}\rangle = \kappa_{V,A} \avg{\bar{q}q}^2,
\end{equation}
where $\kappa_{V,A}$ is a parameter greater than one which accounts for the deviation
from vacuum-state saturation.  In principle, the values of $\kappa$ need not
be identical in the two channels as the four-quark operators are not
identical. In Ref.~\cite{Hohler:2012xd}, it was found that the
QCDSRs in both channels can be satisfied with a single value for $\kappa$.

The vacuum WSRs are constructed from the moments of the difference of the
vector and axial-vector spectral functions, denoted by $\Delta \rho(s) = \rho_V(s)-\rho_A(s)$.
They read
\begin{eqnarray}
&({\rm WSR}\mbox{-}0)& \quad  \int_0^\infty \frac{\Delta \rho(s)}{s^2}
\, \mathrm{d}s = \frac{1}{3} f_{\pi}^2 \avg{r_{\pi}^2} - F_A \ ,
\label{eq:WSR0} \\
&({\rm WSR}\mbox{-}1)& \quad  \int_0^\infty \frac{\Delta \rho(s)}{s} \,
\mathrm{d}s = f_{\pi}^2 \ ,
\label{eq:WSR1} \\
&({\rm WSR}\mbox{-}2)& \quad  \int_0^\infty \Delta \rho(s)  \,
\mathrm{d}s =-2m_q \avg{\bar{q}q} =  f_{\pi}^2 m_{\pi}^2  \ ,
\label{eq:WSR2} \\
&({\rm WSR}\mbox{-}3)& \quad  \int_0^\infty \Delta \rho(s) \, s \,
\mathrm{d}s = -2 \pi \alpha_s \langle\mathcal{O}^{SB}_4\rangle \ .
\label{eq:WSR3}
\end{eqnarray}
Here, $\avg{r_{\pi}^2}$ is the average squared pion radius and $F_A$ is the
radiative pion decay constant.  We have used the Gell-Mann-Oakes-Renner
relation in WSR-2. The chiral order parameter appearing in WSR-3 is the chiral
symmetry-breaking four-quark condensate and can be expressed in terms of the
vector and axial-vector four-quark condensates as
\begin{equation}
\avg{\mathcal{O}^{SB}_4} = \frac{16}{9} \left( \frac{7}{18} \avg{\mathcal{O}_4^{V}} + \frac{11}{18} \avg{\mathcal{O}_4^{A}} \right).
\end{equation}
This condensate can also be written in a factorized form as
$\avg{\mathcal{O}^{SB}_4} = \frac{16}{9} \kappa \avg{\bar{q}q}^2$ with $\kappa = \frac{7}{18}\kappa_V + \frac{11}{18}\kappa_A$.

%%%%%%%%%%%%%%%%%%%%%%%%%%%%%%%%%%%%%%%%%%%%%%%%%%%%%%
\subsection{Finite temperature QCD sum rules}
\label{ssec:QCDSR}
%%%%%%%%%%%%%%%%%%%%%%%%%%%%%%%%%%%%%%%%%%%%%
At finite temperatures, we will only consider mesons with vanishing
three-momentum in the thermal rest frame.
The finite-temperature QCDSRs then have the same structure as in
Eq.~(\ref{eqn:SR}) but with the condensates developing a temperature
dependence and the appearance of additional
nonscalar condensates~\cite{Hatsuda:1992bv}. The latter can be
characterized by their twist, which is defined as the difference of
an operator's
dimension and its spin. The finite-temperature $c_2$ and $c_3$ terms are then
given by
\begin{eqnarray}
 c_{2}^{V/A}(T) &=& \frac{1}{24}{\avg{\frac{\alpha_{s}}{\pi}G^{2}}}_{T} + m_{q} {\avg{\bar{q}q}}_{T}\nonumber\\&&\qquad\qquad\qquad\qquad + \avg{\mathcal{O}^{\tau=2,s=2}}_T, \\
 c_3^V(T)  &=&  -\frac{56}{81} \pi \alpha_s \avg{ \mathcal{O}_4^V }_T + \avg{\mathcal{O}^{\tau=2,s=4}}\nonumber\\&&\qquad\qquad\qquad\qquad+\avg{\mathcal{O}^{\tau=4,s=2}}_T , \\
 c_3^A(T)  &=&  \frac{88}{81} \pi \alpha_s \avg{ \mathcal{O}_4^A }_T + \avg{\mathcal{O}^{\tau=2,s=4}}_T \nonumber\\ && \qquad\qquad\qquad\qquad+\avg{\mathcal{O}^{\tau=4,s=2}}_T,
 \end{eqnarray}
where $\avg{\mathcal{O}^{\tau=2,s=2}}_T$, $\avg{\mathcal{O}^{\tau=2,s=4}}_T$, and
$\avg{\mathcal{O}^{\tau=4,s=2}}_T$ are the twist-2 spin-2, twist-2 spin-4, and
twist-4 spin-2 condensates, respectively. A full
discussion of these and other non-scalar condensates can be found, e.g., in
Refs.~\cite{Hatsuda:1992bv,Leupold:1998bt}.

For a consistent implementation of the chirally mixed spectral functions,
we need to obtain the leading-order temperature dependence of the condensates.
To do so, we must introduce a small parameter about which we are expanding.
In the current setup, there are three scales, $m_\pi$, $T$, and
$\Lambda_{\chi}$, where the latter corresponds to a typical hadronic energy
scale (also representing the condensates). At low temperature and small pion
mass, there are thus two small ratios available,
$m_\pi/\Lambda_{\chi}$ and $T/\Lambda_{\chi}$. For the temperatures in which we
are interested, $m_\pi \sim T$ are of similar size. We define a
parameter
\begin{equation}
\lambda \sim \frac{m_\pi}{\Lambda_{\chi}}, \frac{T}{\Lambda_{\chi}},
\end{equation}
to represent both ratios. Thereby we will work to leading order in $\lambda$.

The temperature dependence of a generic operator, $\mathcal{O}$, can be expressed
in terms of its vacuum value plus its expectation value within external
hadronic states. The matrix elements with single-pion states provide the
dominant contribution at low temperatures; thus the temperature dependence can
be written as
\begin{equation} \label{eq:opT}
\langle \mathcal{O} \rangle_T \simeq \langle \mathcal{O}\rangle_0
+ 3 \int \frac{d^3 k}{\left(2 \pi\right)^3 2 E_k} f^\pi(E_k;T) \
\langle \pi^a(\vec{k})|\mathcal{O}|\pi^a(\vec{k})\rangle
\end{equation}
where $f^\pi$ denotes the thermal Bose distribution and
$E_k = \sqrt{k_{\pi}^{2} + m_{\pi}^{2}}$ the pion energy. Soft pion theorems can
then be applied so that $\langle \pi^a(\vec{k})|\mathcal{O}|\pi^a(\vec{k})\rangle$
becomes momentum independent for scalar operators~\cite{Hatsuda:1992bv}\footnote{Nonscalar
operators can still depend on the pion momentum because of their spin structure.
However, the dependence is usually simple enough that the integral over momentum
can still be performed.}. Corrections to the soft pion theorems are suppressed
by at least one power of $\lambda$. Therefore the entire temperature dependence is
contained in the quantity (adopting standard notation)
\begin{equation}
\label{eq:eps}
\epsilon(T) = \frac{2}{f_{\pi}^{2}} \int_{0}^{\infty} \frac{1}{(2\pi)^{3}}
\frac{\mathrm{d}^{3}p}{E_p} \ f^\pi(E_p;T) \ .
\end{equation}
The function $\epsilon(T)$ is proportional to the scalar pion density, such that
$\epsilon \propto \rho_\pi^s / f_\pi^2$ represents a small parameter. Formally it
is $\mathcal{O}(\lambda^2)$, thus the leading finite temperature effects to the
condensates
are of this order. In the chiral limit, $m_{\pi} =0$, it simplifies to
$\epsilon(T) = T^2/6f_{\pi}^2$; however, for finite pion mass,
we should evaluate the full expression in Eq.~(\ref{eq:eps}).

To find the leading temperature dependence of an operator thus amounts to
evaluating the matrix element $\langle \pi^a|\mathcal{O}|\pi^a\rangle$.
For quark and gluon condensates to order $\lambda^2$
it was derived in Ref.~\cite{Hatsuda:1992bv},
\begin{eqnarray}
{\avg{\bar{q}q}}_{T} &=& {\avg{\bar{q}q}}_0
\left( 1- \frac{3}{4} \epsilon(T) \right), \label{eq:qqT}
\\
 {\avg{\frac{\alpha_{s}}{\pi}G^{2}_{\mu\nu}}}_{T} &=&
{\avg{\frac{\alpha_{s}}{\pi}G^{2}_{\mu\nu}}}_0 - \frac{2}{3} m_{\pi}^2
f_{\pi}^2 \epsilon(T) \ .
\label{eq:g2T}
\end{eqnarray}
The leading temperature dependence of the quark condensate has also been
derived in Ref.~\cite{Gerber:1988tt}, starting from a calculation of the pressure
of a low-temperature gas of finite-mass pions to order $T^8$.
The contribution from single-pion states then determines the leading-order
temperature dependence of the quark and gluon condensates, which we have verified
to match
those found in Ref.~\cite{Hatsuda:1992bv} (note that for finite pion mass, the
leading $T$ dependence of the gluon condensate enters at $T^4$, while for
massless pions it comes in at $T^8$).
The temperature dependence of the four-quark condensates was also found
in Ref.~\cite{Hatsuda:1992bv} for $\kappa=1$. Upon expressing their vacuum
values with $\kappa_{V,A} \neq 1$, one obtains
\begin{eqnarray}
 \label{eqn:SF-T}
 \avg{ \mathcal{O}_4^V }_T  &=& \kappa_{V} \avg{\bar{q}q}^2_0 \left( 1 - \frac{18}{7} \frac{\kappa}{\kappa_V} \epsilon(T) \right), \label{eq:a2T} \\
 \avg{ \mathcal{O}_4^A }_T  &=& \kappa_{A} \avg{\bar{q}q}^2_0 \left( 1 - \frac{18}{11} \frac{\kappa}{\kappa_A} \epsilon(T) \right).
\end{eqnarray}
We note~\cite{Hatsuda:1992bv,Eletsky:1992xd,Kapusta:1994,Leupold:2006ih}
that factorization of the 4-quark condensate is not expected to hold at finite
temperature; the appearance of $\kappa_{V/A}$ in the above expressions is a mere
relic of the parametrization in the vacuum.
The temperature dependence of the nonscalar operators (again to order
$\lambda^2$) is given by~\cite{Hatsuda:1992bv,Leupold:1998bt,Zschocke:2002mn}
\begin{eqnarray}
\avg{\mathcal{O}^{\tau=2,s=2}}_T
&=& A_2^\pi \left(\frac{3}{4} m_{\pi}^{2} I_1(T) + I_2(T)\right) \ ,
\\
\avg{\mathcal{O}^{\tau=2,s=4}}_T
&=& -A_4^\pi \left(\frac{5}{8} m_{\pi}^4 I_1(T)\right.
\nonumber\\
&&\quad \left.+ \frac{5}{2} m_\pi^2 I_2(T) +2 I_3(T)\right) \ ,
\\
\avg{\mathcal{O}^{\tau=4,s=2}}_T
&=& - B_2^\pi \left(\frac{3}{4} m_{\pi}^{2} I_1(T) + I_2(T)\right) \ ,
\end{eqnarray}
where
\begin{equation}
I_n(T) = \int_0^\infty \frac{\mathrm{d}^{3}p}{(2\pi)^3}
\frac{(p^2)^{n-1}}{E_p} \ f^\pi(E_p;T) \ .
\end{equation}
Note that $I_1(T) = f_\pi^2 \epsilon(T)/2$.
The quantities $A_2^\pi$ and $A_4^\pi$ represent moments of the
light-quark distribution functions, $\psi(x,\mu^2)$, inside the pion as
\begin{equation}
A_n = 2 \int_0^1 dx \,x^{n-1} \left[ \psi(x,\mu^2) + \bar{\psi}(x,\mu^2)\right].
\end{equation}
Following Refs.~\cite{Hatsuda:1992bv,Zschocke:2002ic}, we will use
$A_2^\pi = 0.97$ and $A_4^\pi = 0.255$.
The parameter $B_2^\pi$ can
be determined from deep inelastic scattering, which in the case of the nucleon
gives $B_2^N = -0.247 {\rm GeV}^2$.
A similar measurement for the pion is not available. Therefore we will proceed
by taking $B_2^\pi=0$ and assessing deviations from this value as part of
the uncertainty in our calculations.

Since we will need to numerically evaluate the QCDSRs, a metric is needed to
quantify the agreement of the two sides of a given QCDSR.
We use the method proposed by Leinweber~\cite{Leinweber:1995fn} which constructs
the average deviation between the left-hand side (LHS) and right-hand side (RHS)
of the sum rule over a range of $\M^2$\footnote{In this definition, the continuum
contribution to the spectral functions is moved to the RHS of each sum
rule.}. This measure, known as the $d$ value~\cite{Leinweber:1995fn,Leupold:1997dg},
is defined by
\begin{equation}
\label{eq:dvalue}
d \equiv \frac{1}{\Delta\M^2} \int_{\M^2_{\mathrm{min}}}^{\M^2_{\mathrm{max}}}  \left\lvert 1-\frac{\mathrm{LHS}}{\mathrm{RHS}} \right\rvert \, \mathrm{d}\M^2.
\end{equation}
The Borel-mass window, within which the sum rule is expected to be valid, is bound
by $\M^2_{\mathrm{max}}$ and $\M^2_{\mathrm{min}}$. The size of the window is
given by $\Delta\M^2 = \M_{\rm max}^2 - \M^2_{\rm min}$. The lower end of the
window, $\M^2_{\rm min}$, is determined such that the contribution of the
$c_3$ term in the QCDSR has reached 10\% of the OPE side.
The upper end, $\M^2_{\rm max}$, is found so that the continuum
contribution to the LHS of the QCDSR has reached half of the contribution from
the resonances. This definition is slightly different than what has been
used in Refs.~\cite{Leinweber:1995fn,Leupold:1997dg} due to the larger continuum
threshold in the spectral functions used here, but the resulting Borel
window is comparable in size to those of the previous works.

%%%%%%%%%%%%%%%%%%%%%%%%%%%%%%%%%%%%%%%%%%%%%%%%%%%%%%%%%%%
\subsection{Finite temperature Weinberg-type sum rules}
\label{ssec:WSR}
%%%%%%%%%%%%%%%%%%%%%%%%%%%%%%%%%%%%%%%%%%%%%%%%%%%%%%%%%%
The extension of the WSRs to finite temperatures was systematically
investigated in Ref.~\cite{Kapusta:1994}. Corresponding expressions for
WSR-1 and -2 were obtained while also deriving both the vacuum and
finite-temperature
forms of WSR-3.  We are unaware of any finite-temperature analogue to WSR-0.
For mesons at rest in the heat-bath frame (i.e., for vanishing three-momentum),
the in-medium WSRs simplify to
\begin{eqnarray}
\label{eq:WSR1T}
\int_0^\infty \frac{\Delta \bar{\rho}(s,T)}{s} \, \mathrm{d}s &=& 0, \\ \label{eq:WSR2T}
\int_0^\infty \Delta \bar{\rho}(s,T)  \, \mathrm{d}s &=& 0, \\ \label{eq:WSR3T}
\int_0^\infty \Delta \bar{\rho}(s,T) \, s \,  \mathrm{d}s &=& -2 \pi \alpha_s  \avg{\mathcal{O}_4^{SB}}_T .
\end{eqnarray}
As above, we use the notation $\Delta \bar{\rho} \equiv \rho_V - \bar{\rho}_A$
to include the pion pole in the axial-vector spectral
function.\footnote{The vacuum sum rules can also be expressed in terms
of $\bar{\rho}_A$, yielding the same structure as
Eqs.~(\ref{eq:WSR1T})-(\ref{eq:WSR3T})
but with $\avg{\mathcal{O}_4^{SB}}$ evaluated in vacuum.}  This is to
account for the pion developing a nontrivial self-energy in medium.
The temperature dependence of $\avg{\mathcal{O}_4^{SB}}$ was established
in Refs.~\cite{Kapusta:1994,Eletsky:1992xd} as
\begin{equation}
\label{eq:4qXSBT}
\avg{\mathcal{O}_4^{SB}}_T = \avg{\mathcal{O}_4^{SB}}_0 \left[1-2 \epsilon(T)\right].
\end{equation}

%%%%%%%%%%%%%%%%%%%%%%%%%%%%%%%%%
\section{Finite-Temperature Spectral Functions}
\label{sec:FTSF}
%%%%%%%%%%%%%%%%%%%%%%%%%%%%%%%%%%

A main ingredient to the present work is the model-independent leading-order
temperature result for vector and axial-vector spectral functions. In
Ref.~\cite{Dey:1990ba} it was shown that in a dilute pion gas a vector
current interacts with thermal pions to produce an axial-vector current
and vice versa. The in-medium vector and axial-vector spectral
functions then become linear combinations of the vacuum spectral functions,
\begin{eqnarray}
\label{eqn:epsilonmixing1}
\rho_{V}(s,T) &=& \left[1-\epsilon(T)\right]\, \rho_{V}(s,0)
+ \epsilon(T)\, \bar{\rho}_{A}(s,0) \ ,  \\
\label{eqn:epsilonmixing2}
\bar{\rho}_{A}(s,T) &=& \left[1-\epsilon(T)\right]\, \bar{\rho}_{A}(s,0)
+ \epsilon(T)\, \rho_{V}(s,0) \ .
\end{eqnarray}
One can derive these relations in the same way as the temperature dependence of the
operators were determined through an expression analogous to Eq.~(\ref{eq:opT}).
Corrections are therefore likewise suppressed by $\lambda$.
The temperature dependence is also determined by the same parameter
$\epsilon(T)$ given by Eq.~(\ref{eq:eps}).
In the context of the spectral functions, $\epsilon(T)$ takes on the role of a
``mixing parameter" between the vector and axial-vector channels.
Extrapolating the mixing prescription to high temperatures yields degenerate
spectral functions for $\epsilon = 1/2$. Since the above expressions
are based on low-energy effective theory (current algebra), it is implied
that one should mix only the non-perturbative regions of the spectral
functions and leave the perturbative contributions unaltered.

Since the above mixing procedure only relies on vacuum spectral functions,
the agreement of those with vacuum sum rules is essential. Here we employ the
vector and axial-vector spectral functions of Ref.~\cite{Hohler:2012xd}, which
each contain a ground-state resonance, an excited resonance, and a smooth
degenerate continuum. These spectral functions were constructed
to agree with ALEPH $\tau$-decay data~\cite{Barate:1998uf} and further
constrained by the WSRs. Near-perfect agreement was achieved for WSR-1 and -2,
while WSR-0 deviated by $-1.28\%$ and WSR-3 by $-96\%$ (the latter is actually
rather small on the scale of the oscillations
over the nonperturbative integration regime). The negative sign in the
deviations indicates the axial-vector channel's contribution to be larger.
In most previous data-based analyses of the
WSRs~\cite{Barate:1998uf,Ackerstaff:1998yj,Ioffe:2001bn}, the energy integration
was terminated at the $\tau$ mass.  In Ref.~\cite{Hohler:2012xd} a
convergence of the sum rules was established by going beyond this limit, which
dictated the introduction of an $a_1'(1800)$ state as one of the novel features
in that work.
In addition, agreement with the vacuum QCDSRs was found by using
$\alpha_s({\rm 1}$ $\rm{GeV}^2) = 0.5$ and optimizing parameter
values to find $\langle \frac{\alpha _s}{\pi}  G^2_{\mu \nu}  \rangle = 0.022$
${\rm GeV}^4$ and $\kappa_V = \kappa_A = 2.1$, resulting in deviations of
$d_V = 0.24\%$ and $d_A = 0.56\%$
over the Borel windows of $0.085$ ${\rm GeV} < \M < 1.47$ ${\rm GeV}$ for the
vector and $0.089$ ${\rm GeV} < \M < 1.48$ ${\rm GeV}$ for the
axial-vector channel, respectively.\footnote{The value of $\alpha_s$
was chosen in accordance with previous works and to match the energy
scale of the vacuum Borel window, which is centered around 1\,GeV$^2$.
Smaller values of $\alpha_s$ were explored, all yielding significantly
larger $d$ values. Our best-fit value for the gluon
condensate (verified independently of Ref.~\cite{Hohler:2012xd}) turns out to be
somewhat larger than the values inferred in Ref.~\cite{Ioffe:2010zz}, but is not
incompatible with other recent works in the literature, see, e.g., the
discussion in Ref.~\cite{Narison:2011xe}.}

\begin{figure}[!tb]
\centering
\subfigure[Vector Channel]{
\includegraphics[width=.45\textwidth]{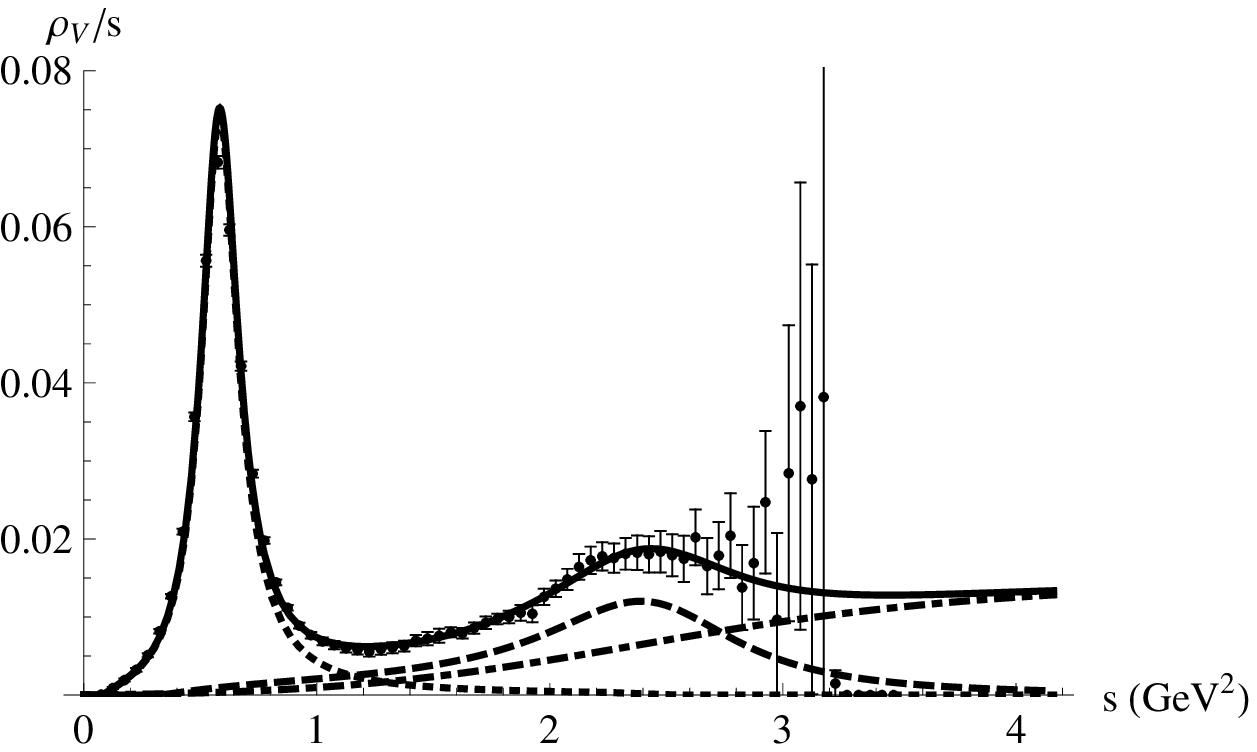}}
\subfigure[Axial-Vector Channel]{
\includegraphics[width=.45\textwidth]{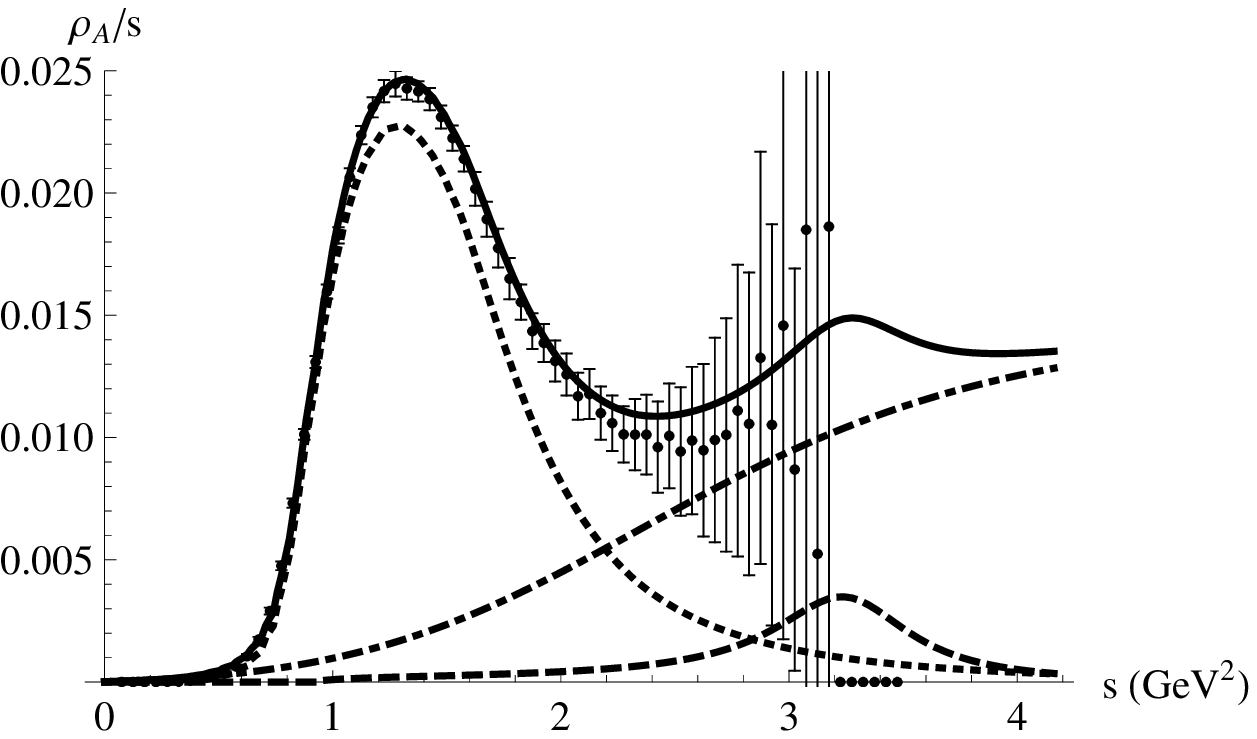}}
\caption{Vacuum spectral functions from Ref.~\cite{Hohler:2012xd} compared with
ALEPH $\tau$-decay data~\cite{Barate:1998uf}. The contributions to the total
spectral function (solid lines) are due to the ground-state resonance
(dotted curve), excited resonance (dashed curve) and a universal
continuum (dotted-dashed curve).}
\label{fig:vacspec}
\end{figure}
The resulting vacuum spectral functions, displayed in Fig.~\ref{fig:vacspec},
contain several key features. The inclusion of excited
states in both channels moves the continuum to higher energies while maintaining (or even improving)
agreement with the $\tau$-decay data. The different masses of the excited vector
and axial-vector states ($\rho'$ and $a_1'$) imply chiral symmetry to be still
broken in this energy region. Consequently, the perturbative regime commences
at higher energies with degenerate continua in the two channels~\cite{Hohler:2012xd},
thus consistently delineating perturbative and nonperturbative components.

\begin{figure}[!tb]
\centering
\subfigure[Vector Channel]{
\includegraphics[width=.45\textwidth]{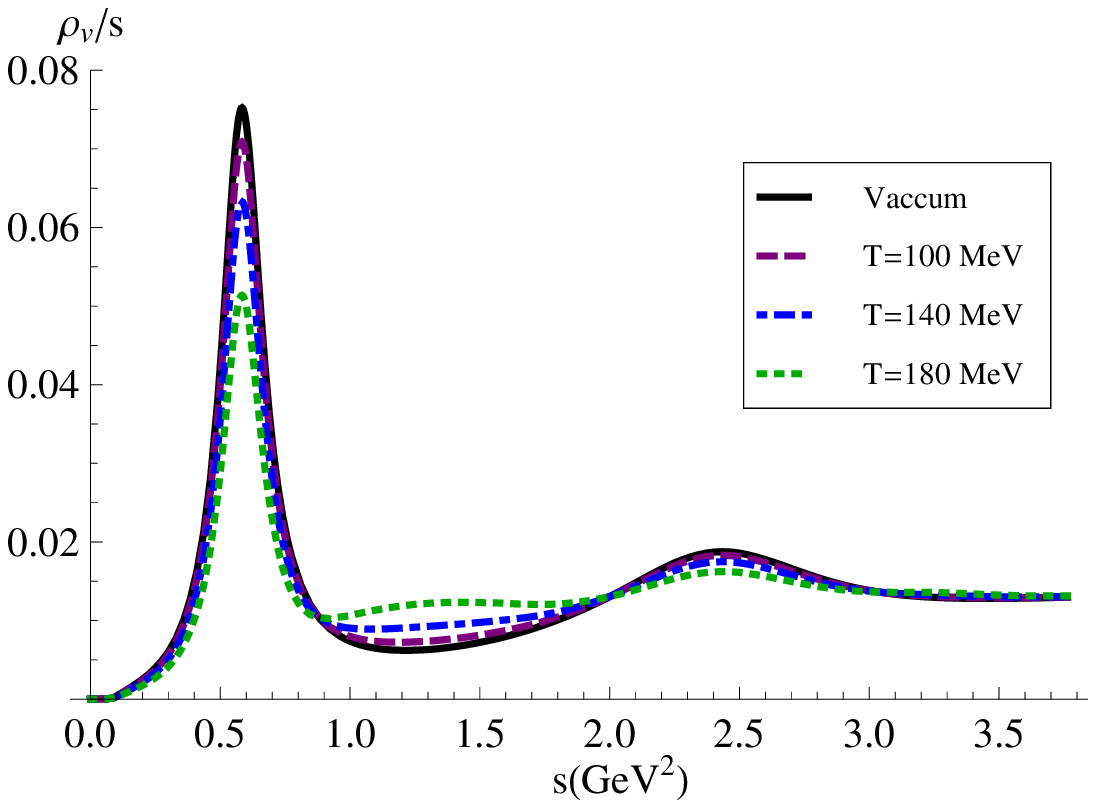}}
\subfigure[Axial-Vector Channel]{
\includegraphics[width=.45\textwidth]{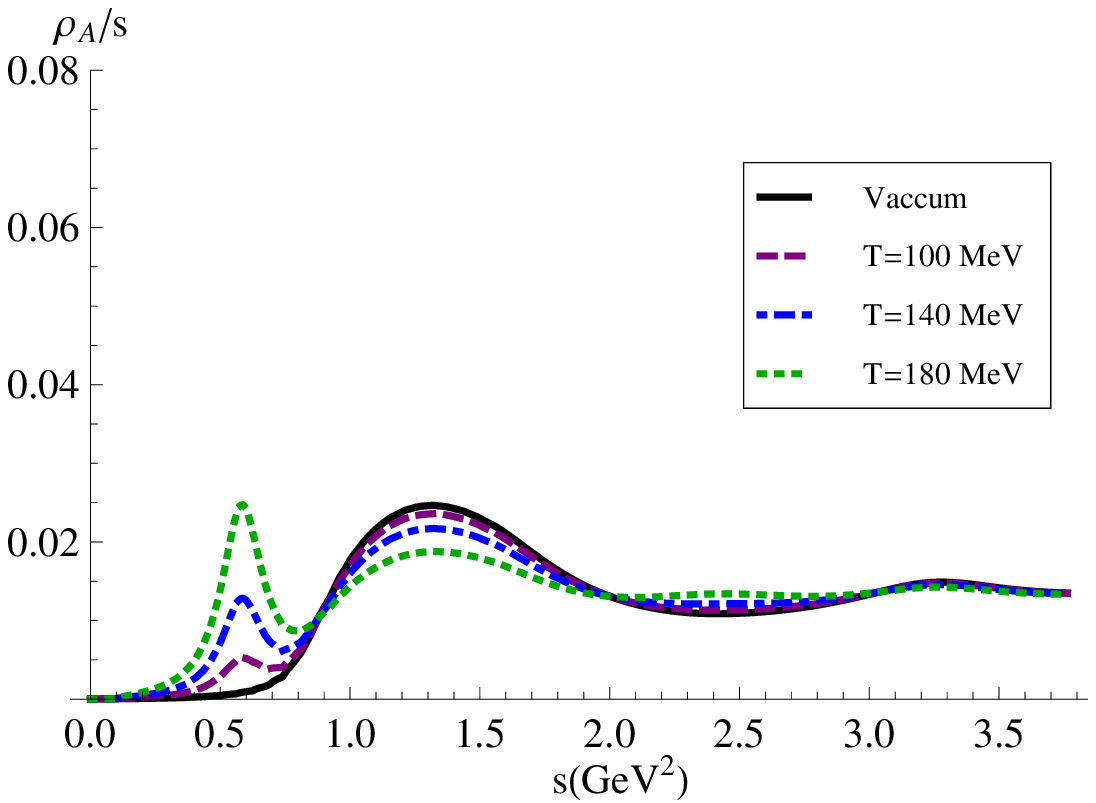}}
\caption{(Finite-temperature vector and axial-vector spectral
functions as following from the chiral mixing of the vacuum ones
of Fig.~\ref{fig:vacspec}.}
  \label{fig:specT}
\end{figure}
The leading-order finite-temperature spectral functions now follow by mixing
the nonperturbative parts of the vacuum spectral functions via
Eqs.~(\ref{eqn:epsilonmixing1}) and (\ref{eqn:epsilonmixing2}),
cf.~Fig.~\ref{fig:specT}. Note that there is no issue of mixing continua
with different thresholds, or perturbative with nonperturbative parts.
Chiral mixing induces a mutual flattening of the
oscillatory pattern of ``peaks" and ``valleys" in the vacuum spectral
functions, with the peaks in one channel filling in the
valleys in the opposite channel: the valley between the $\rho$ and
$\rho'$ is filled by the $a_1$ peak while the valley between
the $a_1$ and $a_1'$ is filled by the $\rho'$ peak, while the spectral strength in the peaks is suppressed.

%%%%%%%%%%%%%%%%%%%%%%%%%%%%%%%%%%%%%%
\section{In-Medium Sum-Rule Analysis}
\label{sec:analysis}
%%%%%%%%%%%%%%%%%%%%%%%%%%%%%%%%%%%%%%%
In this section, we implement the finite-temperature
spectral functions derived from chiral mixing into the WSRs
(Sec.~\ref{ssec:wsr-med}) and QCDSRs (Sec.~\ref{ssec:qcd-med}), followed
by a brief discussion (Sec.~\ref{ssec:disc}).

%%%%%%%%%%%%%%%%%%%%%%%%%%%%%%%%%%%%%%
\subsection{Weinberg sum rules}
\label{ssec:wsr-med}
%%%%%%%%%%%%%%%%%%%%%%%%%%%%%%%%%%%%%%%
The in-medium WSRs, Eqs.~(\ref{eq:WSR1T})-(\ref{eq:WSR3T}), depend on the
difference of the vector and axial-vector spectral functions, $\Delta \bar{\rho}$.
Using chiral mixing, Eqs.~(\ref{eqn:epsilonmixing1}) and
(\ref{eqn:epsilonmixing2}), one has
\begin{equation}
\begin{split}
\Delta \bar{\rho}(s,T) &=\rho_V(s,T) - \bar{\rho}_A(s,T)\\&
= \left[ 1-2\epsilon(T) \right] \Delta \bar{\rho}(s,0) \ .
\end{split}
\end{equation}
Thus the in-medium WSR-1 and -2 remain intact provided their vacuum limit is
satisfied (in fact, residual deviations in the vacuum are suppressed
toward chiral restoration). Since the leading-temperature dependence of the
four-quark condensate is given by ($1-2\epsilon$), WSR-3 is also satisfied
by the chirally mixed spectral functions.
This result holds for any vacuum spectral function provided that no
further adjustments are
applied, e.g., in the treatment of the continua~\cite{Kapusta:1994}.

%%%%%%%%%%%%%%%%%%%%%%%%%%%%%%%%%%%%%%
\subsection{QCD sum rules}
\label{ssec:qcd-med}
%%%%%%%%%%%%%%%%%%%%%%%%%%%%%%%%%%%%%%%

Let us begin by revisiting the QCDSRs in the massless-pion limit of
the chiral-mixing scenario.
One has $\epsilon = T^2/6 f_\pi^2$ so that the leading
temperature dependence is of order $T^2$. Both sides of the sum rule can
then be expanded to this order. The LHS takes the form
\begin{equation}
\begin{split}
&\frac{1}{\M^2}\int \frac{\rho_V(s,0)}{s} e^{-s/\M^2}\\ &\quad\quad + \epsilon(T) \frac{1}{\M^2}\int\frac{\bar{\rho}_A(s,0) - \rho_V(s,0)}{s} e^{-s/\M^2}.
\end{split}
\end{equation}
The second integral can be evaluated using the vacuum QCDSRs.
Because it reflects the difference between the axial-vector and vector
spectral functions, the only term that survives is the chirally breaking
4-quark condensate; thus
\begin{equation}
{\rm LHS} = \frac{1}{\M^2}\int \frac{\rho_V(s,0)}{s} e^{-s/\M^2} + \frac{\epsilon(T)}{\M^6} \pi \alpha_s \avg{\mathcal{O}_4^{SB}}.
\end{equation}
For the RHS (OPE side), the vanishing pion mass makes the quark-condensate term
and the temperature correction to the gluon condensate vanish,
while the non-scalar terms are higher powers of $T^2$ compared with
$\epsilon(T)$. The terms that remain are the $c_0$ term, the gluon condensate,
and the vector 4-quark condensate along with its temperature dependence.
Upon expanding the $c_3$ term, the temperature corrections to the vector
4-quark condensate can be written in terms of the chirally
breaking 4-quark condensate,
\begin{equation}
-\frac{56}{81}\pi \alpha_s \avg{\mathcal{O}_4^V}_T = -\frac{56}{81} \pi \alpha_s \avg{\mathcal{O}_4^V} + \epsilon(T) \pi \alpha_s \avg{\mathcal{O}_4^{SB}}.
\end{equation}
Thus the RHS takes the form
\begin{equation}
\begin{split}
{\rm RHS} &= c_0 + \frac{1}{24 \M^4}\avg{\frac{\alpha_s}{\pi}G_{\mu\nu}^2}\\& + \frac{1}{\M^6}\left(-\frac{56}{81} \pi\alpha_s \avg{\mathcal{O}_4^V} +\epsilon(T)\pi \alpha_s \avg{\mathcal{O}_4^{SB}}\right).
\end{split}
\end{equation}
Comparing the LHS and RHS of the QCDSRs one now sees that the
chirally breaking 4-quark condensate cancels between the two sides while
the remaining terms are simply the vector QCDSR in vacuum. A similar derivation
can be made for the axial-vector channel as well. Thus, for massless
pions, the QCDSRs are analytically satisfied to order $\epsilon$ (see also
Ref.~\cite{Marco:2001dh}).

However, for finite pion mass, some of the simplifications
on the OPE side no longer occur. The expressions for the LHS remain
unchanged, while the temperature dependence of the vector 4-quark condensate
still can be expressed in turns of the chirally breaking 4-quark condensate
so that the cancelation between the two sides for this term occurs. However,
terms which vanished or were temperature suppressed are now present, involving
factors of the squared pion mass.
Most notably, there is an incomplete cancelation between the temperature corrections
of the quark condensate, the gluon condensate, and the twist-2, spin-2 condensates.
A similar lack of cancelation applies to the dimension-6 nonscalar
operators. The balance of these partial cancelations produces terms on the RHS
which are order $\lambda^2$ where there are no such terms on the LHS. Thus
the QCDSRs are explicitly violated at order $\lambda^2$. Therefore we must
proceed by numerically evaluating the QCDSRs to determine
to what degree quantitative deviations affect the validity of the mixing scheme
at a given temperature.

\begin{table*}[htb]
\begin{center}
\begin{tabular}{|c|c|c|c|c|c|c|c|c|c|c|}
\hline
T (MeV)&0&100&110&120&130&140&150&160&170&180\\
\hline
$\epsilon$ & 0 &0.06&0.08&0.10&0.13&0.16&0.20&0.23&0.28&0.32\\
$d_V (\%)$ &0.24&0.32(0.29)&0.38(0.33)&0.48(0.39)&0.64(0.51)&0.85(0.64)&1.11(0.74)&1.43(0.97)&1.82(1.17)&2.29(1.39)\\
$d_A (\%)$ &0.56&0.65(0.57)&0.70(0.58)&0.78(0.61)&0.90(0.67)&1.08(0.76)&1.30(0.88)&1.60(1.01)&1.98(1.17)&2.53(1.34)\\
\hline
\end{tabular}
\end{center}
\caption{Average deviation of the QCDSRs over the Borel window for the
vector and axial-vector channels at select temperatures. Values in parentheses
are based on a frozen Borel window identical to the vacuum one.}
\label{tab:QCDSRT}
\end{table*}

\begin{figure*}[!tbh]
\centering
\subfigure[Vector Channel]{
\includegraphics[width=.24\textwidth]{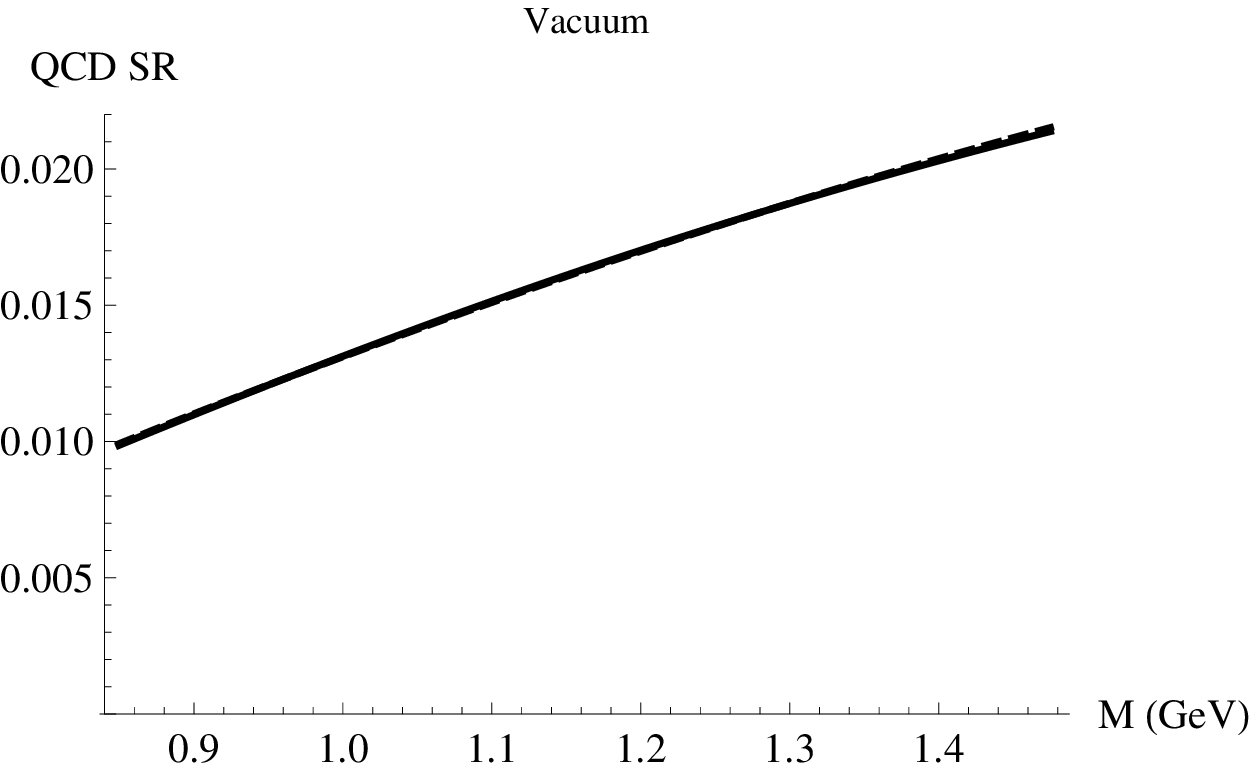}
\includegraphics[width=.24\textwidth]{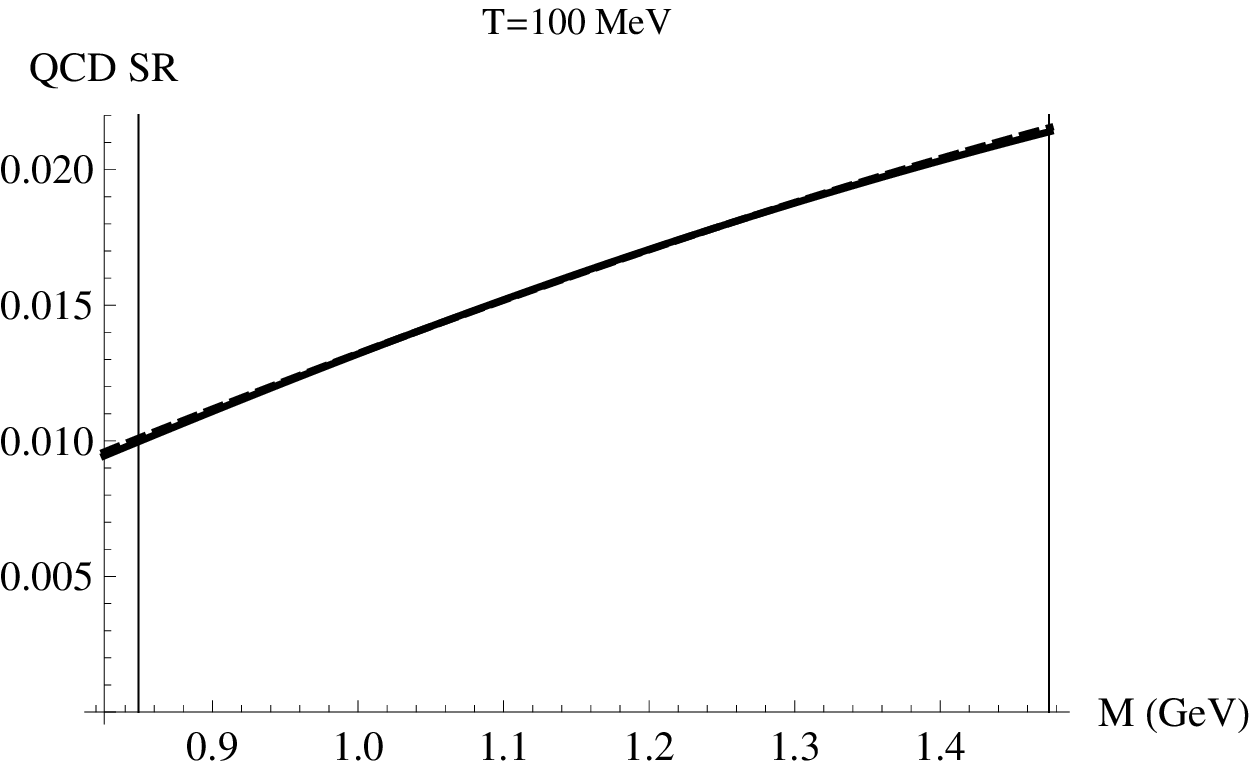}
\includegraphics[width=.24\textwidth]{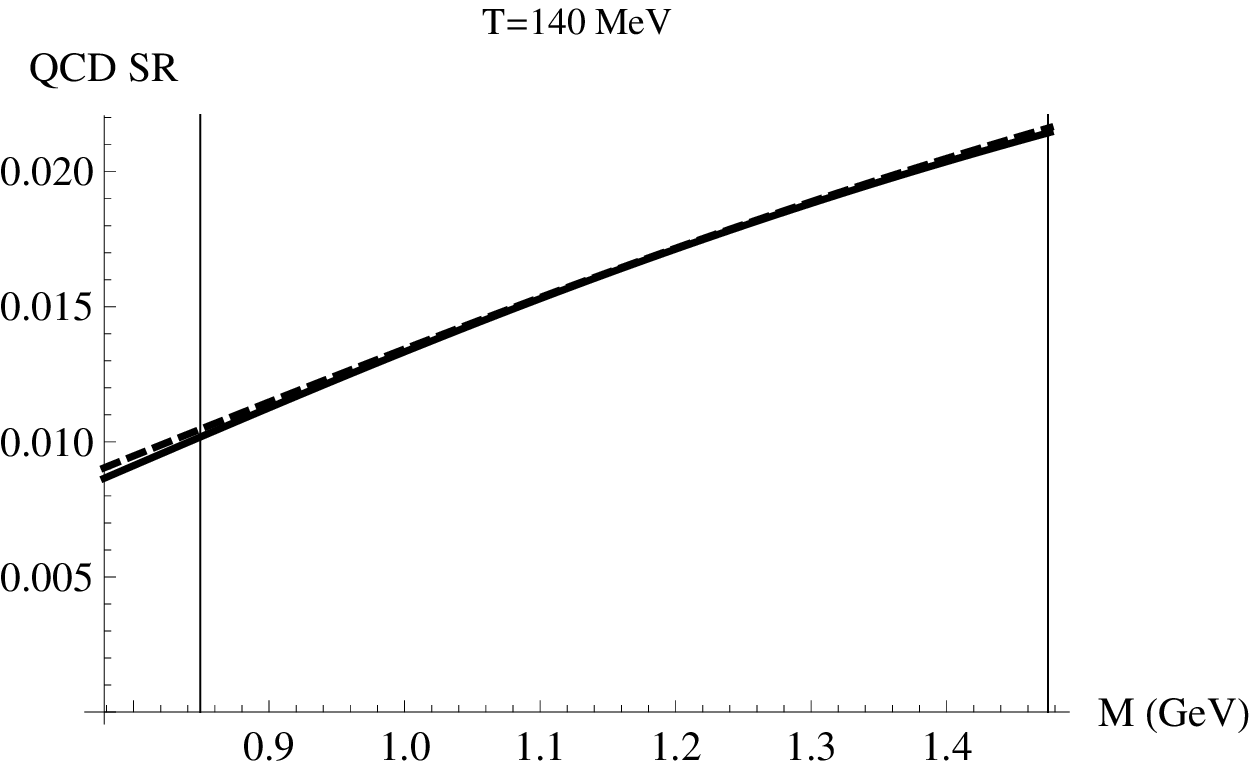}
\includegraphics[width=.24\textwidth]{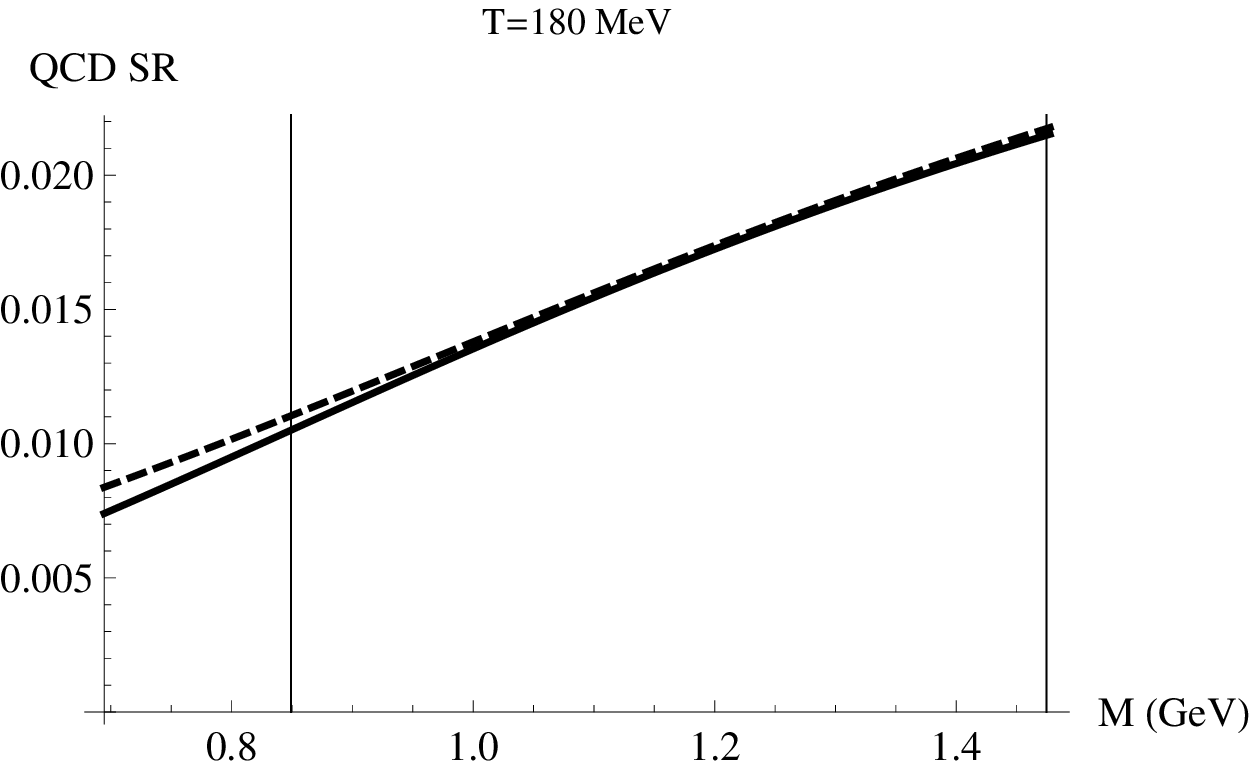}}
\subfigure[Axial-Vector Channel]{
\includegraphics[width=.24\textwidth]{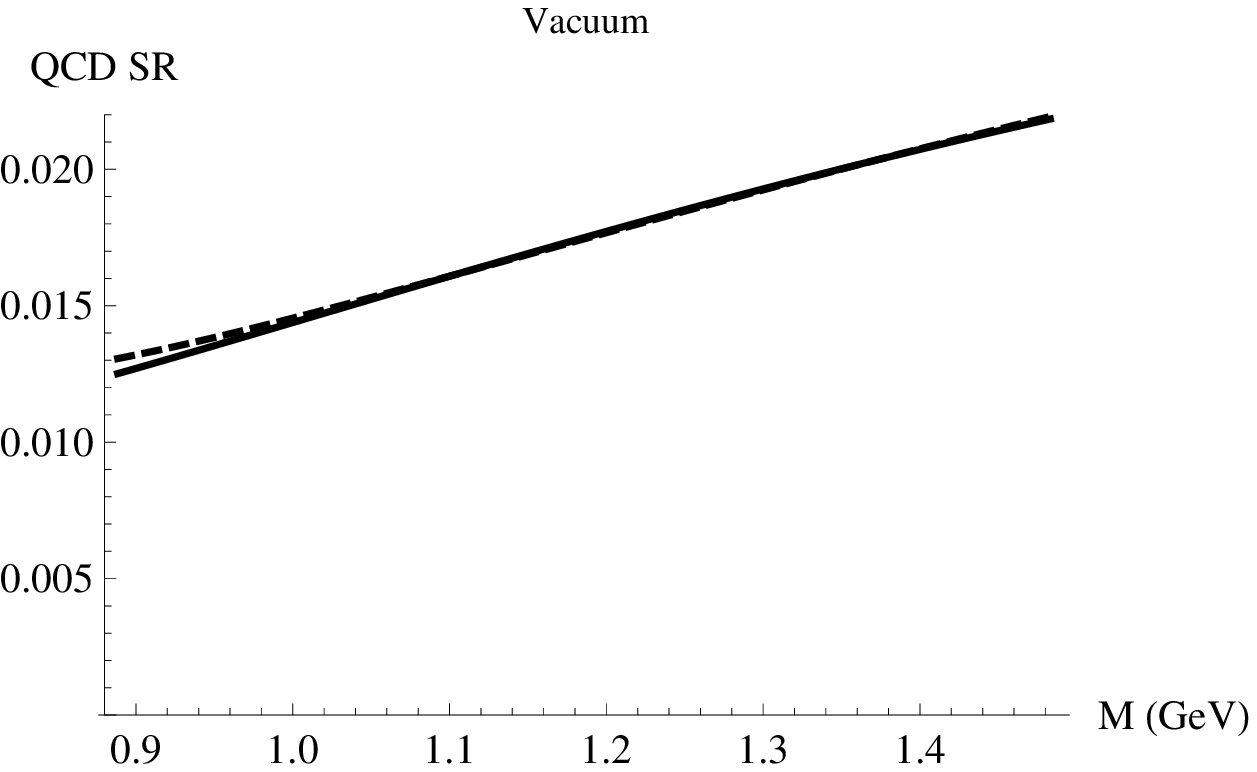}
\includegraphics[width=.24\textwidth]{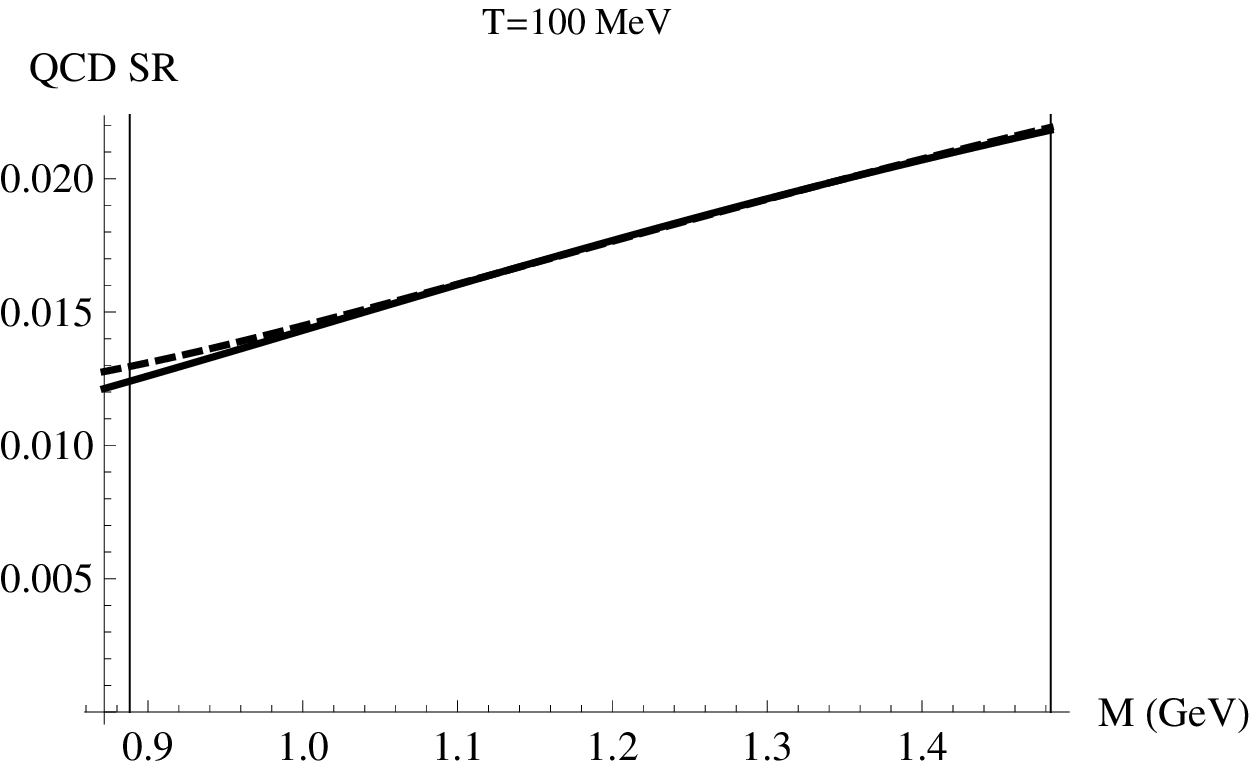}
\includegraphics[width=.24\textwidth]{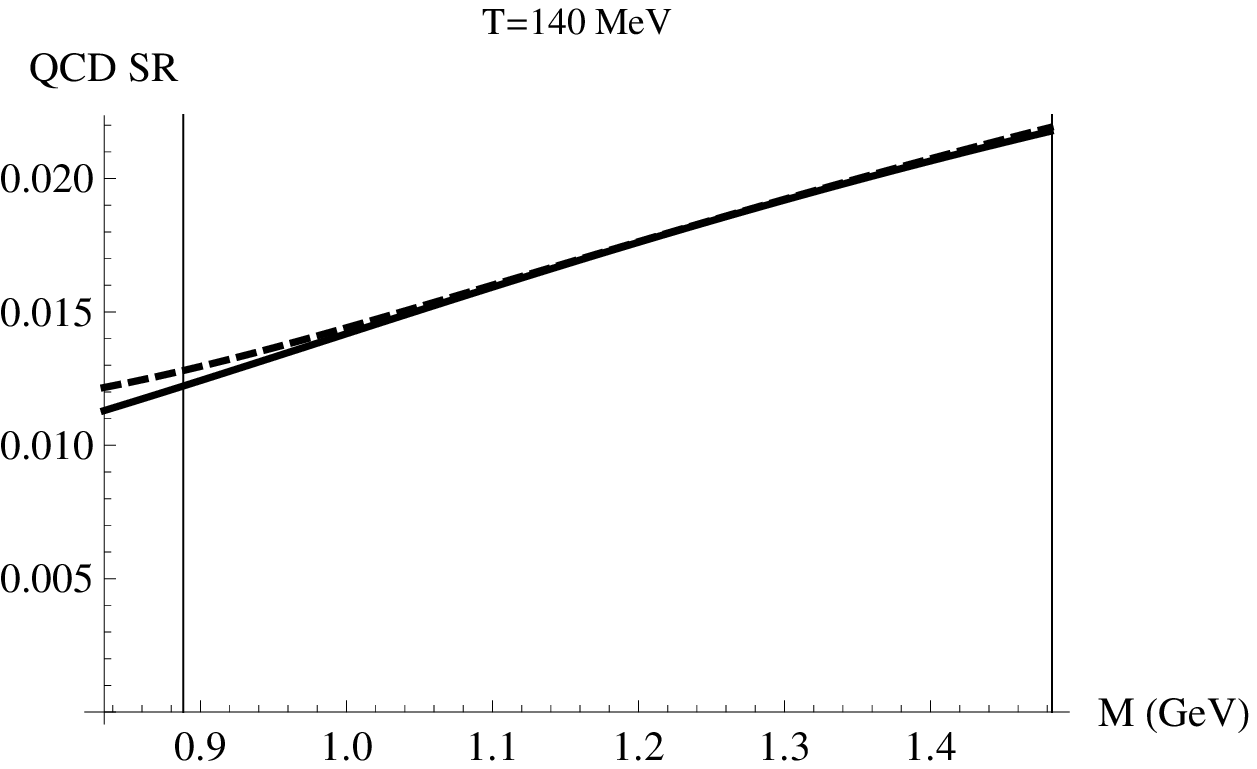}
\includegraphics[width=.24\textwidth]{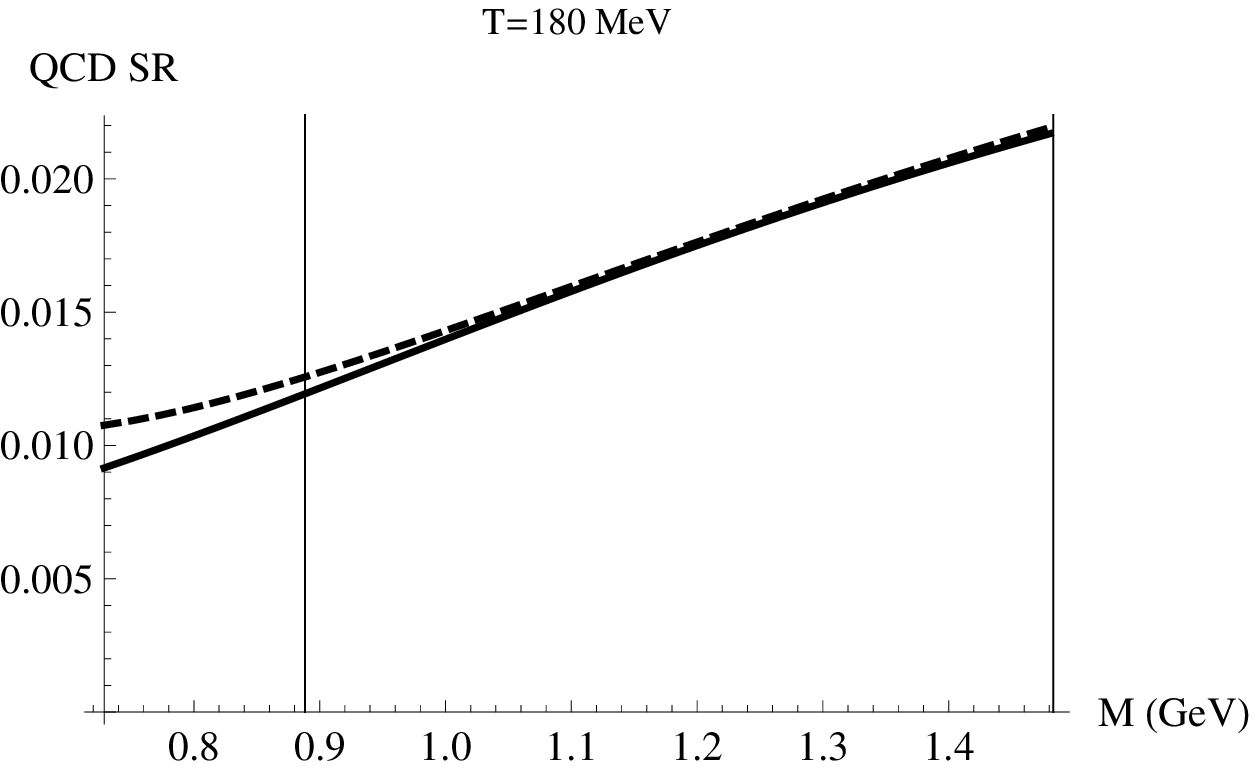}}
\caption{Comparison of the LHS (solid curve) and RHS (dashed curve) of the QCDSRs the vector (upper panels) and axial-vector (lower panels) channels at select temperatures. The extent
of each plot corresponds to the Borel window at that temperature, while the
vertical lines designate the Borel window in vacuum.}
  \label{fig:QCDFT}
\end{figure*}

The two sides of each channel's sum rule are displayed in Fig.~\ref{fig:QCDFT}
for four different
temperatures as a function of Borel mass, while Table~\ref{tab:QCDSRT}
quantifies the average deviation, Eq.~(\ref{eq:dvalue}), over the pertinent in-medium
Borel window. Since neither channel's vacuum spectral functions exactly satisfy the
vacuum QCDSRs, residual vacuum deviations will contribute to the in-medium analysis.
Therefore it is important to compare the in-medium deviations with the vacuum's
deviation.  Overall, the $d$ values stay within a factor of two of the vacuum values for
temperatures up to 120-140\,MeV for the vector and axial-vector channels,
respectively. These temperatures also roughly reflect the range within which
the $d$ values are approximately linear in $\epsilon$, as one would expect in
the regime of validity. Thus one may
conclude that the QCDSRs are reasonably well satisfied up to these temperatures.

Let us address some of the uncertainties in the evaluation of the QCDSRs.
Closer inspection of the Borel window reveals its lower end to expand
significantly with temperature, see Fig.~\ref{fig:QCDFT}. The decrease of
$\M^2_{\rm min}(T)$ can be attributed to the decrease of the $c_3$ term with $T$,
i.e., the 10\%-level contribution of the $1/\M^6$ moves to lower $\M^2$. In
fact, the low-end portion of the Borel window
(extending outside the vacuum window) causes most of the increase of the
deviation at higher temperature. If instead one postulates the in-medium Borel
window to be the same as in vacuum (indicated by the vertical lines in
Fig.~\ref{fig:QCDFT}), the $d$ values decrease appreciably at higher
temperatures (quoted in parenthesis in Table~\ref{tab:QCDSRT}), thus
staying within a 1\% margin up to temperatures as high as 160\,MeV.

Further uncertainties are due to the non-scalar terms in the in-medium
OPE. The accuracy of the twist-2 contributions, $A_2^\pi$ and $A_4^\pi$, is not well
known, while $B_2^\pi$ has not been measured for the pion. We find the impact
of the twist-2 terms on the $d$ values to be small, while $B_2^\pi$ produces a more
significant deviation. For example, a 20\% change in the value
of $A_2^\pi$ causes a {\em relative} change of less than 0.7\% in $d_V$ and $d_A$ for all temperatures.
A similar variation in $A_4^\pi$ causes a larger relative deviation in $d_V$ and $d_A$ of
$\sim 5\%$ at the higher temperatures, but less than $0.1\%$ at the lowest temperature.
A larger uncertainty arises from the numerical value of $B_2^\pi$.
While this parameter has not been experimentally measured for the pion, a
relative estimate of its size may be made by adopting
the value for the nucleon, $B_2^N = -0.247 {\rm GeV}^2$. We then find that $d_V$
is altered by less than 1\% at 100\,MeV but as much as 25\% at 180\,MeV.
The effect on the axial-vector channel is not as pronounced with the maximum
relative deviation of 5\% at the highest temperature.

%%%%%%%%%%%%%%%%%%%%%%%%%%%%%%%%%%%%%%
\subsection{Discussion}
\label{ssec:disc}
%%%%%%%%%%%%%%%%%%%%%%%%%%%%%%%%%%%%%%%
Let us finally put our results into context with previous works on in-medium
sum rules. Most QCDSR analyses thus far have been based on a ground-state
plus continuum
ansatz for the vector and/or axial-vector spectral function. This usually implied
that (a) the continua in the two channels are different, and (b) a reduction of
the in-medium threshold was deduced from the QCDSRs. Both properties can potentially
cause complications when carried out within a rigorous low-temperature
implementation as given by chiral mixing, due to overlaps of the nonperturbative
resonance regime with the perturbative continuum in the two channels. This becomes
particularly problematic in the WSRs, whose low-temperature versions require a
strict mixing of the entire vacuum spectral functions. In the present work, we
resolve this issue by employing vacuum spectral functions with a universal
separation of nonperturbative and perturbative regimes, thanks to excited states
and a degenerate continuum. This ensures a straightforward fulfillment of the
low-temperature WSRs (i.e., within chiral mixing) but also maintains a numerical
agreement of the QCDSRs in both vector and axial-vector channels at low temperature.

The WSRs and the QCDSRs for chirally mixed axial-/vector spectral functions
are analytically satisfied in the chiral limit, without an apparent limitation
due to temperature for their validity.
However, for finite pion mass, the QCDSR are violated at order $\lambda^2$.
The then-required numerical evaluation of the QCDSRs does exhibit deviations
suggesting their breakdown at temperatures of the order of the pion mass.
It is interesting that this breakdown roughly sets in at temperatures where
one expects corrections from higher resonances to become important.
This is not seen if only massless pions are considered.

%%%%%%%%%%%%%%%%%%%%%%%%%
\section{Conclusions}
\label{sec:conclusion}
%%%%%%%%%%%%%%%%%%%%%%%%%
In the present work we have performed a simultaneous analysis of QCD and
Weinberg-type sum rules with vector and axial-vector spectral functions
in the low-temperature limit. Specifically, we have utilized updated vacuum
spectral functions as an ingredient to their chiral mixing, and numerically
evaluated the role of pion mass corrections in the sum rules. The in-medium
spectral functions confirm earlier findings of a mutual flattening trend
with increasing temperature.  While the WSRs could be shown to remain valid
analytically (thereby avoiding a mixing of perturbative and nonperturbative
components in the spectral functions), the finite pion mass corrections
required a numerical analysis of the QCDSRs. They were found to be reasonably
satisfied up to temperatures of about 140\,MeV, with deviations markedly
increasing beyond. This may be an indirect indication for additional
physics, beyond low-energy chiral pion dynamics, which is, of course, well
known from other contexts (resonance excitations).
Our work may be extended to include more realistic medium effects, either
by working to higher orders in $T/\Lambda_{\chi}$ and $m_\pi/\Lambda_{\chi}$,
or by considering
spectral functions which are known to agree with dilepton emission data. It is
instructive to note that even the most simple, yet rigorous, construction
of low-temperature vector and axial-vector spectral functions satisfies sum rules
rooted in QCD while suggesting a mechanism that could play a role toward
their degeneration.

\acknowledgments
The work of N.P.M.H., P.M.H., and R.R. is supported by the US-NSF Grant
No.~PHY-0969394 and by the A.-v.-Humboldt Foundation (Germany).
N.P.M.H. thanks M. Causey for valuable discussions.

\begin{flushleft}
%%%%%%%%%%%%%%%%%%%%%%%%%%%%%%%%

\end{flushleft}

\begin{thebibliography}{10}
%%%%%%%%%%%%%%%%%%%%%%%%%%%%%%%%

%\cite{Shuryak:2004book}
\bibitem{Shuryak:2004book}
 E.~V.~Shuryak,
 % ``The QCD Vacuum, Hadrons, and Superdense Matter''
 World Sci.\ Lect.\ Notes Phys.\ {\bf 71}, 1 (2004).

%\cite{Ioffe:2010zz}
\bibitem{Ioffe:2010zz}
  B.L.~Ioffe, V.S.~Fadin, and L.N.~Lipatov,
  \textit{Quantum Chromodynamics: Perturbative and Nonperturbative Aspects}
  (Cambridge University Press, New York, 2010)

%\cite{Friman:2011zz}
\bibitem{Friman:2011zz}
  Edited by B.~Friman, C.~Hohne, J.~Knoll, S.~Leupold, J.~Randrup, R.~Rapp, and P.~Senger,
  %``The CBM physics book: Compressed baryonic matter in laboratory experiments''
  Lect.\ Notes Phys.\  {\bf 814}, 1 (2011).
  %%CITATION = LNPHA,814,1;%%


 %\cite{Satz:2012zz}
\bibitem{Satz:2012zz}
  H.~Satz,
 %``Extreme states of matter in strong interaction physics. An introduction,''
  Lect.\ Notes Phys.\  {\bf 841}, 1 (2012).
  %%CITATION = LNPHA,841,1;%%
%%%%%%%%%%%%%%%%%%%%%%%%%%%%%%%%%%%%%%%%%%%
  %\cite{PDG:2012}
\bibitem{PDG:2012}
  J. Beringer \textit{et al.} (Particle Data Group), Phys. Rev. D{\bf 86}, 010001 (2012).


%\cite{Borsanyi:2010bp}
\bibitem{Borsanyi:2010bp}
  S.~Borsanyi \textit{et al.}  (Wuppertal-Budapest Collaboration),
  %``Is there still any T_c mystery in lattice QCD? Results with physical masses in the continuum limit III,''
  J. High Energy Phys. 09 (2010) 073.
 % [arXiv:1005.3508 [hep-lat]].
  %%CITATION = ARXIV:1005.3508;%%

%\cite{Bazavov:2011nk}
\bibitem{Bazavov:2011nk}
  A.~Bazavov, T.~Bhattacharya, M.~Cheng, C.~DeTar, H.~T.~Ding, S.~Gottlieb, R.~Gupta and P.~Hegde \textit{et al.},
  %``The chiral and deconfinement aspects of the QCD transition,''
 Phys.\ Rev.\ D {\bf 85}, 054503 (2012).
%  [arXiv:1111.1710 [hep-lat]].
  %%CITATION = ARXIV:1111.1710;%%


%\cite{Shifman:1978bx}
\bibitem{Shifman:1978bx}
  M.~A.~Shifman, A.~I.~Vainshtein and V.~I.~Zakharov,
  %``QCD and Resonance Physics. Sum Rules,''
  Nucl.\ Phys.\ {\bf B147}, 385 (1979).
  %%CITATION = NUPHA,B147,385;%%

  %\cite{Shifman:1978by}
\bibitem{Shifman:1978by}
  M.~A.~Shifman, A.~I.~Vainshtein and V.~I.~Zakharov,
  %``QCD and Resonance Physics: Applications,''
  Nucl.\ Phys.\ {\bf B147}, 448 (1979).
  %%CITATION = NUPHA,B147,448;%%

%\cite{Das:1967ek}
\bibitem{Das:1967ek}
  T.~Das, V.~S.~Mathur and S.~Okubo,
  %``Low-energy theorem in the radiative decays of charged pions,''
  Phys.\ Rev.\ Lett.\  {\bf 19}, 859 (1967).
  %%CITATION = PRLTA,19,859;%%

 %\cite{Weinberg:1967kj}
 \bibitem{Weinberg:1967kj}
  S.~Weinberg,
  %``Precise relations between the spectra of vector and axial vector mesons,''
  Phys.\ Rev.\ Lett.\  {\bf 18}, 507 (1967).
  %%CITATION = PRLTA,18,507;%%

%\cite{Kapusta:1994}
\bibitem{Kapusta:1994}
  %``Weinberg-type sum rules at zero and finite temperature,''
  J.~I.~Kapusta and E.~V.~Shuryak,
  Phys.\ Rev.\ D {\bf 49}, 4694 (1994).
%%%%%%%%%%%%%%%%%%%%%%%%%%%%%%%%%%%%%%%%%%%%%%%%%%%%%%%%%%%%

%\cite{Dey:1990ba}
\bibitem{Dey:1990ba}
  M.~Dey, V.~L.~Eletsky and B.~L.~Ioffe,
  %``Mixing of vector and axial mesons at finite temperature: an Indication towards chiral symmetry restoration,''
  Phys.\ Lett.\ B {\bf 252}, 620 (1990).
  %%CITATION = PHLTA,B252,620;%%

%\cite{Steele:1996su}
\bibitem{Steele:1996su}
  J.~V.~Steele, H.~Yamagishi and I.~Zahed,
  %``Dilepton and photon emission rates from a hadronic gas,''
  Phys.\ Lett.\ B {\bf 384}, 255 (1996).
%  [hep-ph/9603290].
  %%CITATION = HEP-PH/9603290;%%

%\cite{Chanfray:1998hr}
\bibitem{Chanfray:1998hr}
  G.~Chanfray, J.~Delorme and M.~Ericson,
  %``Chiral symmetry restoration and parity mixing,''
  Nucl.\ Phys.\ {\bf A637}, 421 (1998).
  %[nucl-th/9801020].
  %%CITATION = NUCL-TH/9801020;%%

%\cite{Krippa:1997ss}
\bibitem{Krippa:1997ss}
  B.~Krippa,
  %``Chiral symmetry and mixing of axial and vector correlators in matter,''
  Phys.\ Lett.\ B {\bf 427}, 13 (1998).
 % [hep-ph/9708365].
  %%CITATION = HEP-PH/9708365;%%

%\cite{Harada:2008hj}
\bibitem{Harada:2008hj}
  M.~Harada, C.~Sasaki and W.~Weise,
  %``Vector-axialvector mixing from a chiral effective field theory at finite temperature,''
  Phys.\ Rev.\ D {\bf 78}, 114003 (2008).
 % [arXiv:0807.1417 [hep-ph]].
  %%CITATION = ARXIV:0807.1417;%%
%%%%%%%%%%%%%%%%%%%%%%%%%%%%%%%%%%%%%%%%%%%%%%%%%%%%%%%%%%%%%%%

  %\cite{Eletsky:1992xd}
\bibitem{Eletsky:1992xd}
  V.~L.~Eletsky,
  %``Four quark condensates at T not = 0,''
  Phys.\ Lett.\ B {\bf 299}, 111 (1993).
  %%CITATION = PHLTA,B299,111;%%

%%%%%%%%%%%%%%%%%%%%%%%%%%%%%%%%%%%%%%%%%%%%%%%%%%%%%%%%%%%%%%%%%

%\cite{Hatsuda:1992bv}
\bibitem{Hatsuda:1992bv}
  T.~Hatsuda, Y.~Koike and S.~-H.~Lee,
  %``Finite temperature QCD sum rules reexamined: rho, omega and A1 mesons,''
  Nucl.\ Phys.\ {\bf B394}, 221 (1993).
  %%CITATION = NUPHA,B394,221;%%


%\cite{Hofmann:1999nn}
\bibitem{Hofmann:1999nn}
  R.~Hofmann, T.~Gutsche and A.~Faessler,
  %``Thermal QCD sum rules in the rho0 channel revisited,''
  Eur.\ Phys.\ J.\ C {\bf 17}, 651 (2000).
 % [hep-ph/9907351].
  %%CITATION = HEP-PH/9907351;%%

%\cite{Zschocke:2002mn}
\bibitem{Zschocke:2002mn}
  S.~Zschocke, O.~P.~Pavlenko and B.~Kampfer,
  %``Evaluation of QCD sum rules for light vector mesons at finite density and temperature,''
  Eur.\ Phys.\ J.\ A {\bf 15}, 529 (2002).
 % [nucl-th/0205057].
  %%CITATION = NUCL-TH/0205057;%%
%%%%%%%%%%%%%%%%%%%%%%%%%%%%%%%%%%%%%%%%%%%%%%%%%%%%%%%%%

%\cite{Marco:2001dh}
\bibitem{Marco:2001dh}
  E.~Marco, R.~Hofmann and W.~Weise,
  %``Note on finite temperature sum rules for vector and axial vector spectral functions,''
  Phys.\ Lett.\ B {\bf 530}, 88 (2002).
 % [hep-ph/0110110].
  %%CITATION = HEP-PH/0110110;%%

%\cite{Kwon:2010fw}
\bibitem{Kwon:2010fw}
  Y.~Kwon, C.~Sasaki and W.~Weise,
  %``Vector mesons at finite temperature and QCD sum rules,''
  Phys.\ Rev.\ C {\bf 81}, 065203 (2010).
 % [arXiv:1004.1059 [nucl-th]].
  %%CITATION = ARXIV:1004.1059;%%


%\cite{Hohler:2012xd}
\bibitem{Hohler:2012xd}
  P.M.~Hohler and R.~Rapp,
  %``Sum rule analysis of vector and axial-vector spectral functions with excited states in vacuum,''
  Nucl.\ Phys.\ {\bf A892}, 58 (2012).
 % [arXiv:1204.6309 [hep-ph]].
  %%CITATION = ARXIV:1204.6309;%%
  
  %ALEPH
%\cite{Barate:1998uf}
\bibitem{Barate:1998uf}
  R.~Barate \textit{et al.} (ALEPH Collaboration),
  %``Measurement of the spectral functions of axial - vector hadronic tau decays and determination of alpha(S)(M**2(tau)),''
  Eur.\ Phys.\ J.\ C {\bf 4}, 409 (1998).
  %%CITATION = EPHJA,C4,409;%%

% OPAL
%\cite{Ackerstaff:1998yj}
\bibitem{Ackerstaff:1998yj}
  K.~Ackerstaff \textit{et al.} (OPAL Collaboration),
  %``Measurement of the strong coupling constant alpha(s) and the vector and axial vector spectral functions in hadronic tau decays,''
  Eur.\ Phys.\ J.\ C {\bf 7}, 571 (1999).
%  [hep-ex/9808019].
  %%CITATION = HEP-EX/9808019;%%

%\cite{Leupold:1998bt}
\bibitem{Leupold:1998bt}
  S.~Leupold and U.~Mosel,
  %``On QCD sum rules for vector mesons in nuclear medium,''
  Phys.\ Rev.\ C {\bf 58}, 2939 (1998).
 % [nucl-th/9805024].
  %%CITATION = NUCL-TH/9805024;%%

%%%%%%%%%%%%%%%%%%%%%%%%%%%%%%%
%\cite{Gerber:1988tt}
\bibitem{Gerber:1988tt}
  P.~Gerber and H.~Leutwyler,
  %``Hadrons Below the Chiral Phase Transition,''
  Nucl.\ Phys.\ {\bf B321}, 387 (1989).
  %%CITATION = NUPHA,B321,387;%%
  %343 citations counted in INSPIRE as of 21 Mar 2013

%\cite{Leupold:2006ih}
\bibitem{Leupold:2006ih}
  S.~Leupold,
  %``Four-quark condensates and chiral symmetry restoration in a resonance gas model,''
  J.\ Phys.\ G {\bf 32}, 2199 (2006).
%  [hep-ph/0604058].
  %%CITATION = HEP-PH/0604058;%%

 %\cite{Zschocke:2002ic}
\bibitem{Zschocke:2002ic}
  S.~Zschocke, B.~Kampfer, O.~P.~Pavlenko and G.~Wolf,
  %``Evaluation of QCD sum rules for HADES,''
  arXiv:nucl-th/0202066.
  %%CITATION = NUCL-TH/0202066;%%

 %\cite{Leinweber:1995fn}
\bibitem{Leinweber:1995fn}
  D.~B.~Leinweber,
  %``QCD sum rules for skeptics,''
  Ann. Phys.\ (NY) {\bf 254}, 328 (1997).
 % [nucl-th/9510051].
  %%CITATION = NUCL-TH/9510051;%%

%\cite{Leupold:1997dg}
\bibitem{Leupold:1997dg}
  S.~Leupold, W.~Peters, and U.~Mosel,
  %``What QCD sum rules tell about the rho meson,''
  Nucl.\ Phys.\ {\bf A628}, 311 (1998).
 % [nucl-th/9708016].
  %%CITATION = NUCL-TH/9708016;%%

\bibitem{Ioffe:2001bn}
  B.~L.~Ioffe,
  %``Chiral effective theory of strong interactions,''
    Usp.\ Fiz.\ Nauk {\bf 171}, 1273 (2001)
   [Phys.\ Usp.\ {\bf 44}, 1211 (2001)].


%\cite{Narison:2011xe}
\bibitem{Narison:2011xe}
  S.~Narison,
  %``Gluon Condensates and m_{c,b} from QCD-Moments and their ratios to Order alpha_s^3 and <G^4>,''
  Phys.\ Lett.\ B {\bf 706}, 412 (2012).





\end{thebibliography}
\end{document}